SOCIAL SCIENCES: Social Sciences

# A High-Resolution Human Contact Network for Infectious Disease Transmission

Marcel Salathé[1*] Maria Kazandjieva[2], Jung Woo Lee[2], Philip Levis[2], Marcus W. Feldman[1] & James H. Jones[3,4]

1. Department of Biology, Stanford University, Stanford, CA, USA
2. Department of Computer Sciences, Stanford University, Stanford, CA, USA
3. Department of Anthropology, Stanford University, Stanford, CA, USA
4. Woods Institute for the Environment, Stanford University, Stanford, CA, USA

***Corresponding author:** Marcel Salathé, Department of Biology, Stanford University, Serra Mall, Gilbert 116, Stanford, CA 94305-5020, Stanford, USA.

Tel: (650) 723-6929; Fax: (650) 725-8244; E-mail: salathe@stanford.edu

Current address:

Center for Infectious Disease Dynamics, Department of Biology, Pennsylvania State University, University Park, PA 16802.

**The most frequent infectious diseases in humans - and those with the highest potential for rapid pandemic spread – are usually transmitted via droplets during close proximity interactions (CPIs). Despite the importance of this transmission route, very little is known about the dynamic patterns of CPIs. Using wireless sensor network technology, we obtained high-resolution data of CPIs during a typical day at an American high school, permitting the reconstruction of the social network relevant for infectious disease transmission. At a 94% coverage, we collected 762,868 CPIs at a maximal distance of 3 meters among 788 individuals. The data revealed a high density network with typical small world properties and a relatively homogenous distribution of both interaction time and interaction partners among subjects. Computer simulations of the spread of an influenza-like disease on the weighted contact graph are in good agreement with absentee data during the most recent influenza season. Analysis of targeted immunization strategies suggested that contact network data are required to design strategies that are significantly more effective than random immunization. Immunization strategies based on contact network data were most effective at high vaccination coverage.**

# INTRODUCTION

Pandemic spread of an infectious disease is one of the biggest threats to society due to the potentially high mortality and high economic costs associated with such an event (1, 2). Understanding the dynamics of infectious disease spread through human communities will facilitate the development of much-needed mitigation strategies (3). Schools are particularly vulnerable to infectious disease spread because of the high frequency of close proximity interactions (CPI) which most infectious disease transmission depends upon (3, 4). Infections that are transmitted predominantly via the droplet route, such as influenza, common colds, whooping cough, SARS-CoV and many others are among the most frequent infectious diseases. Droplets from an infected person can reach a susceptible person in close proximity, typically a distance of less than 3 meters (5, 6), making CPIs highly relevant for disease spread. However, very little is known about the dynamic patterns of CPIs in human communities. Here, we present data collected with a wireless sensor network deployment using TelosB motes (7) to detect high-resolution proximity (up to 3 meters) between subjects in a US high school. To our knowledge, the data set represents the first high-resolution temporal contact network relevant to the spread of infectious diseases via droplet transmission in a school.

Previous attempts to capture the contact networks relevant for infectious disease transmission have been based on data collection using surveys, socio-technological networks, and mobile devices such as cell phones. Each of these approaches has advantages and disadvantages. Surveys manage to capture the interactions relevant for disease transmission, but are often limited by small sample sizes (8) and are subject to human error (9). Socio-technological networks can provide large, long-term data sets (10, 11) but fail to capture the CPIs relevant for disease transmission. The use of mobile devices aware of their location (or of other mobile devices in proximity) represents a promising third alternative. Using mobile phones to detect spatial proximity of subjects is possible with repeated Bluetooth scans (9) but the resolution is too coarse for diseases that are transmitted through the close contact route. Our approach is free of human error, captures the vast majority (94%) of the community of interest and allows us to collect high-resolution contact network data relevant for infectious disease transmission.

Most efforts to understand and mitigate the spread of pandemic diseases (influenza in particular) have made use of large-scale, spatially explicit models parameterized with data from various sources such as census data, traffic / migration data, demographic data etc. (3, 4, 12-15). The population is generally divided into communities of schools, workplaces and households, but detailed data on mixing patterns in such communities are scarce. In particular, very little is known about the contact networks in schools (16) even though schools are known to play a crucially important role in pandemic spread, mainly owing to the intensity of CPIs at schools. In what follows, we describe and analyze the contact network observed at a US high school during a typical school day. Using an SEIR simulation model, we investigate the spread of influenza on the observed contact network and find that the results are in very good agreement with absentee data from the influenza A (H1N1) spread in the fall of 2009. Finally, we implement and test various immunization strategies to evaluate their efficacy in reducing disease spread within the school.

**RESULTS**

The data set covers close proximity interactions (CPIs) of 94% of the entire school population - 655 students, 73 teachers, 55 staff and 5 other - and contains 2,148,991 unique close proximity records (CPR). A CPR represents one close (max. 3 meters) proximity detection event between two motes. An *interaction* is defined as a continuous sequence ($\geq 1$) of CPR between the same two motes, and a *contact* is the sum of all interactions between these two motes. Thus, a contact exists between two motes if there is at least one interaction between them during the day, and the duration of the contact is the total duration of all interactions between these two motes. Because the beaconing frequency of a mote is 0.05 s$^{-1}$, an interaction of length 3 (in CPR) corresponds to an interaction of about one minute (see Supplementary Material and reference therein). The entire data set consists of 762,868 interactions with a mean duration of 2.8 CPR (~ 1 minute), or 118,291 contacts with mean duration of 18.1 CPR (~ 6 minutes). Figure 1a shows the frequency *f(m)* of interactions and contacts of length *m* (in minutes). The majority of interactions and contacts are very short (80$^{th}$ percentile of interactions at 3 CPR; 80$^{th}$ percentile of contacts at 15 CPR), and even though about 80% of the total time is spent in interactions that are shorter than 5 minutes, short contacts (less than 5 minutes) represent only about 10% of the total time (Figure 1b).

The temporal mixing patterns observed are in accordance with the schedule of the school day, i.e., the average degree (number of contacts) peaks between classes and during lunch breaks (Figure 1c). The aggregate network for the entire day can be represented by a weighted, undirected graph where nodes represent individuals and edges represent contacts (edges are weighted by contact duration). The topology of the contact network is an important determinant of infectious disease spread (17, 18). Traditional infectious disease models assume that all subjects have the same number of contacts, or that the contact network of subjects is described by a random graph with a binomial degree distribution. However, many networks from a wide range of applications - including contact networks relevant for infectious disease transmission (19, 20) - have been found to have highly heterogeneous degree distributions. Such heterogeneity is important because it directly affects the basic reproductive number $R_0$, a crucially important indicator of how fast an infectious disease spreads and what fraction of the population will be infected. In particular, if $\rho_0$ is the incorrect estimate for $R_0$ in a heterogeneous network under the false assumption of a uniform degree distribution, the correct estimate is given by $R_0 = \rho_0 \,[1+(CV)^2]$ where $CV$ is the coefficient of variation of the degree (17, 21). Thus, the coefficient of variation quantifies the extent to which contact heterogeneity affects disease dynamics.

The descriptive statistics of the school network with different definitions of contact are shown in Figure 2. In order to account for the fact that the majority of the contacts are relatively short (Figure 1a), we recalculated all statistics of the network with a minimum requirement for contact duration $c_m$ (i.e. all edges with weight $< c_m$ are removed from the graph). The network exhibits typical "small world" properties (22) such as a relatively high transitivity (also known as clustering coefficient, which measures the ratio of triangles to connected triplets) and short average path length for all values of $c_m$. Assortativity, the tendency of nodes to associate with similar nodes with respect to a given property (23), was measured with respect to degree and role of the person (i.e. student, teacher, etc.). Interestingly, while both measures are relatively high, degree assortativity decreases while role assortativity increases with higher values of $c_m$. Due to the very high density of the contact network, a giant component exists for all values of $c_m$. Community structure (or modularity) is relatively high, increasingly so with higher values of $c_m$, indicating that more intense contacts tend to occur more often in subgroups and less often between such groups (24). We find a very homogenous degree distribution with the squared coefficient of variation of the degree distribution $(CV)^2 = 0.118$ for the full network, and slightly increased heterogeneity in the network with higher cutoff values $c_m$ (Figure 2j). The distributions of number of interactions $c$ and the strength $s$, the weighted equivalent of the degree (25) are equally homogenous (Figure 3). Overall, the data suggest that the network topology is best described by a low variance small world network.

To understand infectious disease dynamics at the school, we used an SEIR simulation model (parameterized with data from influenza outbreaks, see Methods for details) where an index case becomes infected outside the school on a random day during the week, and disease transmission at the school occurs during weekdays on the full contact network as described by the collected data. Each individual is chosen as an index case for 1000 simulation runs, resulting in a total of 788,000 epidemic simulation runs. This simulation setting represents a base scenario where a single infectious case introduces the disease into the school population. In reality, multiple introductions are to be expected if a disease spreads through a population, but the base scenario used here allows us to quantify the predictive power of graph-based properties of individuals on epidemic outcomes. We assume that symptomatic individuals remove themselves from the school population after a few hours. We find that in 67.7% of all simulations, no secondary infections occur and thus there is no outbreak, whereas in the remaining 32.3% of the simulations, outbreaks occur with an average attack rate of 3.87% (all simulations: 1.33%; maximum 46.19%), and the average $R_0$, measured as the number of secondary infections caused by the index case, is 3.85 (all simulations: 1.24; maximum 18). Absentee data from the school during the fall of 2009 (i.e. during the second wave of H1N1 influenza in the northern hemisphere) are in good agreement with simulation data generated by the SEIR model running on the contact network (Figure 4a).

A strong correlation exists between the size of an outbreak caused by index case individual $i$ and the strength of the node representing individual $i$ ($r^2 = 0.929$). The correlation between outbreak size and degree is substantially weaker ($r^2 = 0.525$) because at the high temporal resolution of the data set, the degree contains many short duration contacts whose impact on epidemic spread is minimal. To estimate the sampling rate at which degree has maximal predictive power, we systematically sub-sampled our original data set to yield lower resolution data sets. Figure 4b shows that sampling as infrequently as every 100 minutes would have resulted in the same predictive power for degree as sampling every 20 seconds, while the maximum predictive power for degree would have been attained at ~ 20 minutes. At this sampling rate, the 95% confidence intervals for the correlation between degree and outbreak size and the correlation between strength and outbreak size start to overlap (due to the high correlation between degree and strength, blue line in Figure 4a). These results suggest that high resolution sampling of network properties such as the degree of nodes might be highly misleading for prediction purposes if used in isolation (i.e. without the temporal information that allows for weighing).

To mitigate epidemic spread, targeted immunization interventions or social distancing interventions aim to prevent disease transmission from person to person. Finding the best immunization strategy is of great interest if only incomplete immunization is possible, as is often the case at the beginning of the spread of a novel virus. In recent years, the idea of protecting individuals based on their position in the contact network has received considerable attention (11, 26, 27). Graph-based properties such as node degree, node betweenness centrality (28), etc. have been proposed to help identify target nodes for control strategies such as vaccination, but due to the lack of empirical contact data on closed networks relevant for the spread of influenza-like diseases, such strategies could only be tested on purely theoretical networks (or on approximations from other empirical social networks that did not measure CPIs directly (11)). To understand the effect of partial vaccination, we measured outbreak size for three different levels of vaccination coverage (5%, 10% and 20%) and a number of different control strategies based on node degree, node strength, betweenness centrality, closeness centrality and eigenvector centrality (so called "graph-based strategies"). In addition, we tested vaccination strategies that do not require contact network data (random vaccination, preferential vaccination for teachers and preferential vaccination for students, see Methods). In order to ensure robustness of the results to variation in transmission probabilities, all simulations were tested with three different transmission probabilities (see Methods). Figure 4c shows which strategies led to significantly ($p < 0.05$, two-sided Wilcoxon test) different outcomes at all transmission probability values. As expected, all strategies managed to significantly reduce the final size of the epidemic. Compared to the random strategy, graph-based strategies had an effect only at higher vaccination coverage. Graph-based strategies did not differ much in their efficacy; in general, strength-based strategies were the most effective. Overall, two main results emerge: (i) in the absence of information on the contact network, all available strategies, including random immunization, performed equally well. (ii) In the presence of information on the contact network, high resolution data supports a strength-based strategy, but there was no significant difference among the graph-based strategies.

**DISCUSSION**

In summary, we present high resolution data from the close proximity interaction network at a US high school during a typical school day. Notably, the month of the experiment (January) is associated with the second highest percentage of influenza cases in the US for the 1976–77 through 2008–09 influenza seasons (second only to February). The data suggest that the network relevant for disease transmission is best described as a small world network with very homogenous contact structure, in which short, repeated interactions dominate. The low values of the coefficients of variation in degree, strength and number of interactions (Figure 3) suggest that the assumption of homogeneity in traditional disease models (21) might be sufficiently realistic for simulating the spread of influenza-like diseases in communities such as high schools. Furthermore, we do not find any "fat tails" in the contact distribution of our data set, corroborating the notion (8) that the current focus on networks with such distributions is not warranted for infectious disease spread within local communities.

It is important to recognize the limitations of the data presented here, particularly in light of the fact that transmission of influenza-like diseases also occurs via other routes, for example via contact with contaminated surfaces (29). Moreover, different pathogens as well as different strains of a particular pathogen might have different minimum requirements (both spatial and temporal) that need to be met for person-to-person transmission. At present, the data capture the contact network during a single day only. However, this is not an inherent short-coming of the approach presented here, and while there is no obvious reason to assume that the large scale structure of the contact network in a high school would change substantially from day to day, long-term studies could address this issue in the future. Data collection at different schools with different demographic compositions would be helpful in clarifying if and how demographic compositions affect the properties of the network relevant for disease transmission. Wireless sensor network technology certainly allows further elucidation of the contact networks not only at different schools, but also in households, hospitals, workplaces and other community settings.

With regard to immunization strategies, our simulation results suggest that contact network data is necessary to design strategies that are significantly more effective than random immunization in order to minimize the number of cases at the school caused by a single index case. Great care needs to be taken in interpreting these results, for various reasons. First, the limitations of the data as discussed above mean that these results may not hold in other settings, underlining the need for further empirical network studies. Second, the simulations assume neither multiple introductions nor ongoing interactions of participants outside of the school. To what extent these assumptions, particularly the latter, are violated when a disease spreads through a community is unknown and remains to be measured. Third, and perhaps most importantly, a particular immunization strategy may be optimal for reducing the number of cases in one particular school, but may at the same time not be optimal from the perspective of an entire community. Immunization strategies must also take into account medical, social and ethical aspects (30). Thus, while we believe that data of the kind reported here can help inform public health decisions, in particular as more data become available in the future, it is clear that at this stage one cannot derive public health recommendations directly from this study. We note, however, that our findings support the notion that graph-based immunization strategies could in principle help mitigate disease outbreaks (11, 27).

## METHODS

On January 14$^{th}$ 2010, we distributed wireless senor network motes (TelosB (7)) to all students, teachers and staff at an American high school (the date was chosen because it represented a typical school day). Participants were asked to sign an assent form on which they also indicated at what time the mote was turned on. The assent form also asked participants to indicate their role/status at the school, with the following four options available: "student", "teacher", "staff", and "other". At the end of the day, we collected the motes and assent forms, and obtained data with the corresponding assent from 789 motes / individuals. Some of the motes had not been used (due to people either being absent from the school, or not participating in the project), and for some motes with data we did not obtain written assent to use the data. The remaining data cover 94% of the entire school population. We also deployed motes at fixed locations (stationary motes), but these are not part of the data set described here except for one stationary mote in the main cafeteria – the signal of this mote was used to reconstruct the global timestamp (see below). For technical details of the deployment, see Supplementary Material.

Epidemic Simulations

To simulate the spread of an influenza-like illness (ILI), we used an SEIR simulation model parameterized with data from influenza outbreaks (12, 31, 32). In the following, we describe the model in detail.

Transmission occurs exclusively along the contacts of the graph as collected at the school. Each individual (i.e. node of the network) can be in one of four classes: **S**usceptible, **E**xposed, **I**nfectious, and **R**ecovered. Barring vaccination, all individuals are initially susceptible (see further below for more information on vaccination). At a random time step during the first week of the simulation, an individual is chosen as the index case and its status is set to exposed. A simulation is stopped after the number of both exposed and infectious individuals has gone back to 0, i.e. all infected individuals have recovered. Each time step represents 12 hours and is divided into day time and night time. Transmission can occur only during day time, and only on weekdays (i.e. apart from the initial infection of the index case, we do not consider any transmission outside of the school – while this assumption will not hold in reality, it allows us to focus exclusively on within-school transmission and to analyze the spread of a disease starting from a single infected case).

Transmission of disease from an infectious to a susceptible individual occurs with probability 0.003 per 20 seconds of contact (the interval between two beacons). This value has been chosen because it approximates the time-dependant attack rate observed in an outbreak of influenza aboard a commercial airliner (31). In particular, the probability of transmission per time step (12 hours) from an infectious individual to a susceptible individual is *1 - (1-0.003)$^w$* where *w* is the weight of the contact edge (in CPR). Upon infection, an individual will move into the exposed class (infected but not infectious). After the incubation period, an exposed individual will become symptomatic and move into the infectious class. The incubation period distribution is modeled by a right-shifted Weibull distribution with a fixed offset of half a day (power parameter 2.21 and scale parameter 1.10, see (12)). On the half day that the individual becomes infectious, the duration of all contacts of the infectious individual is reduced by 75%. This reduction ensures that if an individual becomes symptomatic and starts to feel ill during a school day, social contacts are reduced and the individual leaves the school or is dismissed from school after a few hours. In the following days, all contacts are reduced by 100% until recovery (i.e. the individual stays at home). Once an individual is infectious, recovery occurs with a probability of *1-0.95$^t$* per time step, where *t* represents the number of time steps spent in the infectious state (in line with data from an outbreak of H1N1 at a New York City School (32)). After 12 days in the infectious class, an individual will recover if recovery hasn't occured before.

We assume that all exposed individuals developed symptoms. High incidence of asymptomatic spread may affect infectious disease dynamics (33), but reports of asymptomatic individuals excreting high levels of influenza virus are rare (34). In addition, a recent community-based study investigating naturally acquired influenza virus infections found that only 14% of infections with detectable shedding at TR-PCR (reverse-transcription polymerase chain reaction) were asymptomatic, and viral shedding was low in these cases (35), suggesting that the asymptomatic transmission plays a minor role. Similar patterns were observed for SARS-CoV, another virus with the potential for rapid pandemic spread: asymptomatic cases were infrequent, and lack of transmission from asymptomatic cases was observed in several countries with SARS outbreaks (36).

Vaccination

The efficacy of vaccination strategies was tested by simulation. Vaccination occurs (if it occurs at all) before introduction of the disease by the index case. Vaccinated individuals are moved directly into the recovering class. We assume that the vaccine provides full protection during an epidemic.

Three vaccination strategies are implemented that do not require measuring graph-based properties – these strategies are called "random", "students" and "teachers".

**Random**: Individuals are chosen randomly until vaccination coverage is reached.

**Students**: Students only are chosen randomly until vaccination coverage is reached.

**Teachers**: Teachers only are chosen randomly until vaccination coverage is reached If that vaccination coverage is so high that all teachers get vaccinated before the coverage is reached, then the strategy continues as student strategy (see above) for the remaining vaccinations.

Three vaccination strategies are implemented that do require measuring graph properties – these strategies are called "degree", "strength" and "betweenness".

**Degree**: Individuals are ranked according to their degree (i.e. number of contacts during the day of measurement). Individuals are then chosen according to that ranking (in descending order) until vaccination coverage is reached.

**Strength**: Individuals are ranked according to their strength (i.e. total time exposed to others during the day of measurement). Individuals are then chosen according to that ranking (in descending order) until vaccination coverage is reached.

**Betweenness**: Individuals are ranked according to their betweenness centrality (see formula below). Individuals are then chosen according to that ranking (in descending order) until vaccination coverage is reached. Betweenness centrality $C_B(i)$ of individual $i$ is calculated as

$$C_B(i) = \sum_{s \neq t \neq i} \frac{\sigma_{st}(i)}{\sigma_{st}}$$

where $s$, $t$ and $i$ are distinct individuals in the contact graph, $\sigma_{st}$ is the total number of shortest paths between nodes $s$ and $t$, and $\sigma_{st}(i)$ is the number of those shortest paths that go through node $i$ (28). The shortest path is calculated using inverse weights.

**Closeness**: Individuals are ranked according to their closeness centrality (see formula below). Individuals are then chosen according to that ranking (in descending order) until vaccination coverage is reached. Closeness centrality $C_C(i)$ of individual $i$ is calculated as

$$C_C(i) = \frac{n-1}{\sum_{s \neq i} d_{si}}$$

where *s* and *i* are distinct individuals in the contact graph, $d_{si}$ is the shortest path between nodes *s* and *i,* and *n* is the number of individuals in the graph (28). The shortest path is calculated using inverse weights.

**Eigenvector**: Individuals are ranked according to their eigenvector centrality. Calculation of eigenvector centrality is described in (37) through application of the pagerank algorithm with jumping probability 0. The measure captures the fraction of time that a random walk would spend at a given vertex during an infinite amount of time.

We tested three different levels of vaccination coverage: 5%, 10%, and 20%. These percentages apply to the entire population, i.e. a 10% vaccination coverage means that 10% of the entire school population is vaccinated, independent of the particular vaccination strategy (except for strategy "none", which means no vaccinations occur). In addition to the default transmission probability per CPR interval described above (i.e. 0.003), we also tested lower (0.002) and higher (0.0045) transmission probability values.

# ACKNOWLEDGEMENTS

This research was supported by a NSF award (BCS-0947132), a Branco Weiss fellowship to M.S., NICH award 1K01HD051494 to J.H.J., NIH grant GM28016 to M.W.F. We would like to thank Ignacio Cancino, Elena V. Jordan, Alison Brown, Rahel Salathé and members of the Feldman, Levis and Jones groups for help with the mote deployments, Marc Horowitz providing a crucial link and two anonymous referees for their valuable comments. We are particularly grateful to the staff members of the school who made this project possible. Special thanks to the creators and maintainers of the Java Universal Network/Graph Framework (JUNG) and of the R package iGraph.

**FIGURE LEGENDS**

**Figure 1**

**(a)** Normalized frequency *f(m)* of interactions and contacts of duration *m* (in minutes) on a log-log scale and **(b)** percentage *p* of total time of all CPIs by interactions and contacts with a minimum duration $c_m$ (in minutes). Most CPI time is spent in medium duration contacts consisting of repeated, short interactions. **(c)** Temporal dynamics of the average number of contacts (degree). Here, the degree of an individual is measured as the number of other individuals in close proximity during 5 minutes. Gray background spans the 2.5% and 97.5% percentile of the degree distribution.

**Figure 2**

Various statistics on the contact graph with minimum contact duration $c_m$ (i.e. the leftmost point in each panel represents the full contact graph, the rightmost point represents the contact graph that contains only contacts that are at least 60 minutes long). With increasing $c_m$, nodes drop out of the network if they have no contact that satisfies the minimum duration condition (hence the reduction in number *V* of nodes in **(a)**). **(b)** Density of the graph ( $2E / (V*(V-1))$ ) where *E* is the number of edges, **(c)** average degree, **(d)** average strength where strength of a node is the total number of CPR of the node, **(e)** transitivity (i.e. cluster coefficient) as defined in (25) and expected value (mean degree / *V*) in random network (dashed line), **(f)** average path length, **(g)** assortativity (23) with respect to degree (black) and with respect to role (red), **(h)** size of the largest component as a fraction of total network size, **(i)** modularity *Q* as defined in (38), **(j)** square of coefficient of variation of degree, $CV^2$.

**Figure 3**

Distribution and squared coefficient of variation of **(a)** degree $d$, **(b)** number of interactions $c$ and **(c)** strength $s$, based on the full contact network and colored by role of individuals.

**Figure 4**

**(a)** Absentee data (red) and data generated by the SEIR model (gray; 1000 runs with $R_0>1$ shown). Gray lines show frequency of infectious individuals, f(I), red line shows the combined frequency of students who reported - or were diagnosed with - a fever and teachers who were absent (gap in the line due to weekend). **(b)** Correlation ($r^2$) between outbreak size and degree of index case (black), outbreak size and strength of index case (red), and degree and strength of index case (blue) at various sampling rates. The leftmost correlations are based on the full data set (sampling interval 1/3 minute), all others are based on sub-sampled data sets that would have been generated with the given sampling interval. Shaded area behind line shows 95% confidence interval of correlation. **(c)** Differences in effect of vaccination strategies. 10,000 simulations for each combination of vaccination strategy, vaccination coverage and transmission probability (see Methods) with a random index case per simulation were recorded (i.e. the panel represents a total of 810,000 simulations). Colors represent vaccination coverage 5% (orange), 10% (blue) and 20% (gray). A point at the intersection of strategy A and strategy B indicated that between those strategies, there was a significant difference ($p < 0.05$ two-sided Wilcoxon test) in the outbreak size at all transmission probability values at the given vaccination coverage.

A black horizontal or vertical line points in the direction of the strategy that resulted in smaller outbreak sizes. Due to the symmetry of the grid, data points below the left bottom – top right diagonal line are not shown.

# FIGURES

**Figure 1**

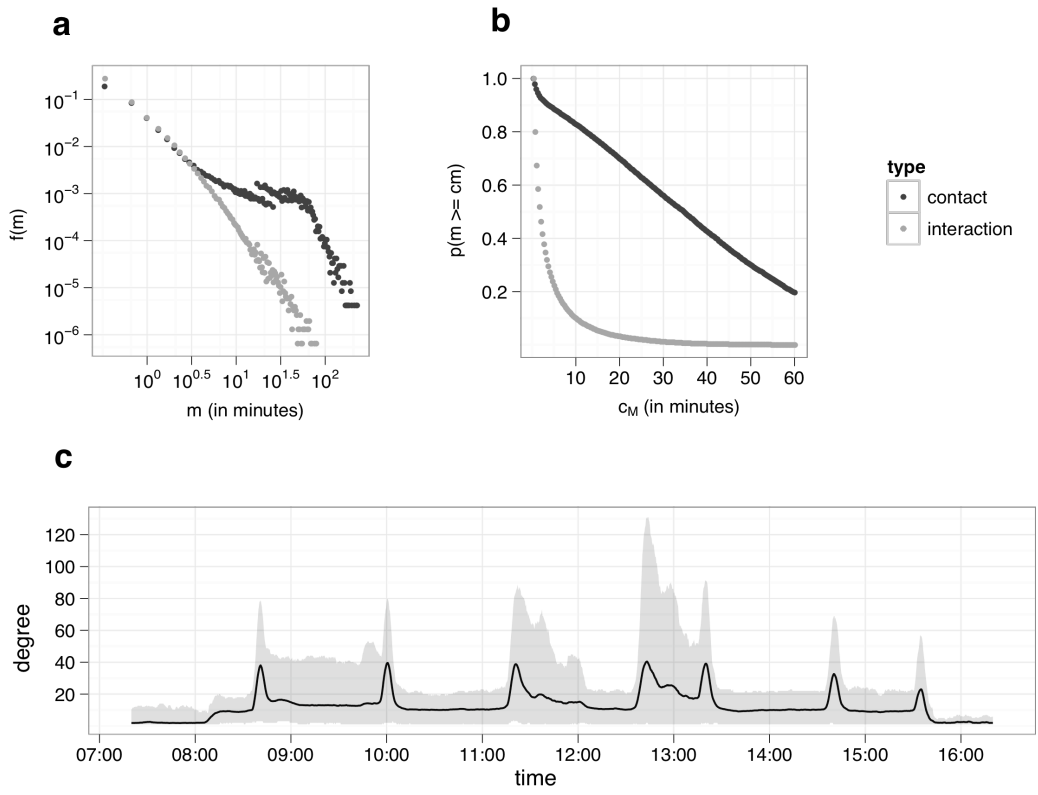

**Figure 2**

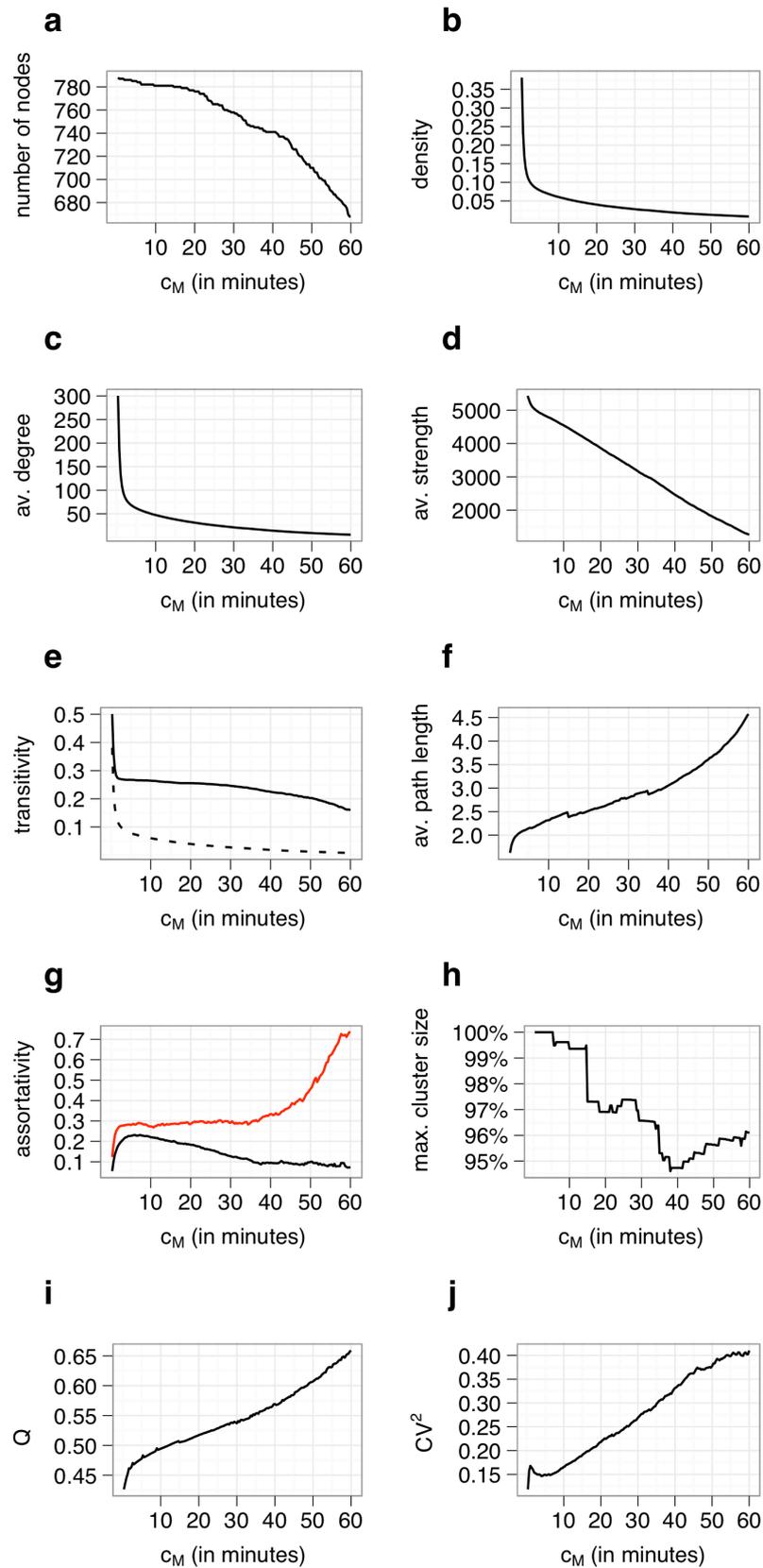

**Figure 3**

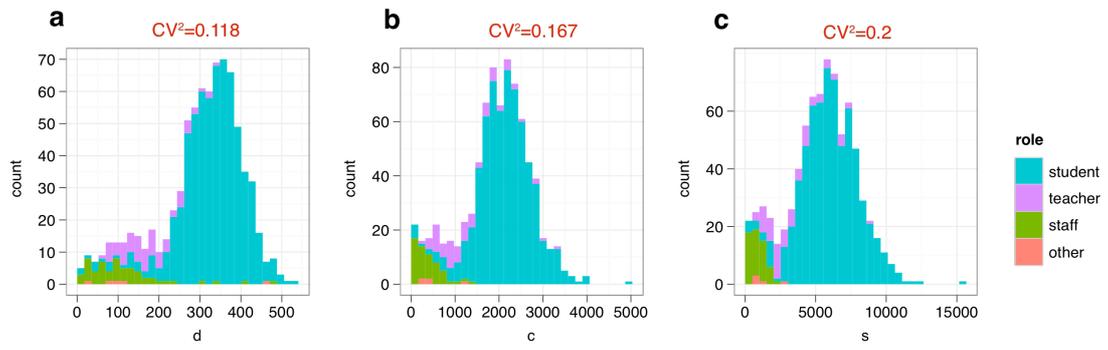

**Figure 4**

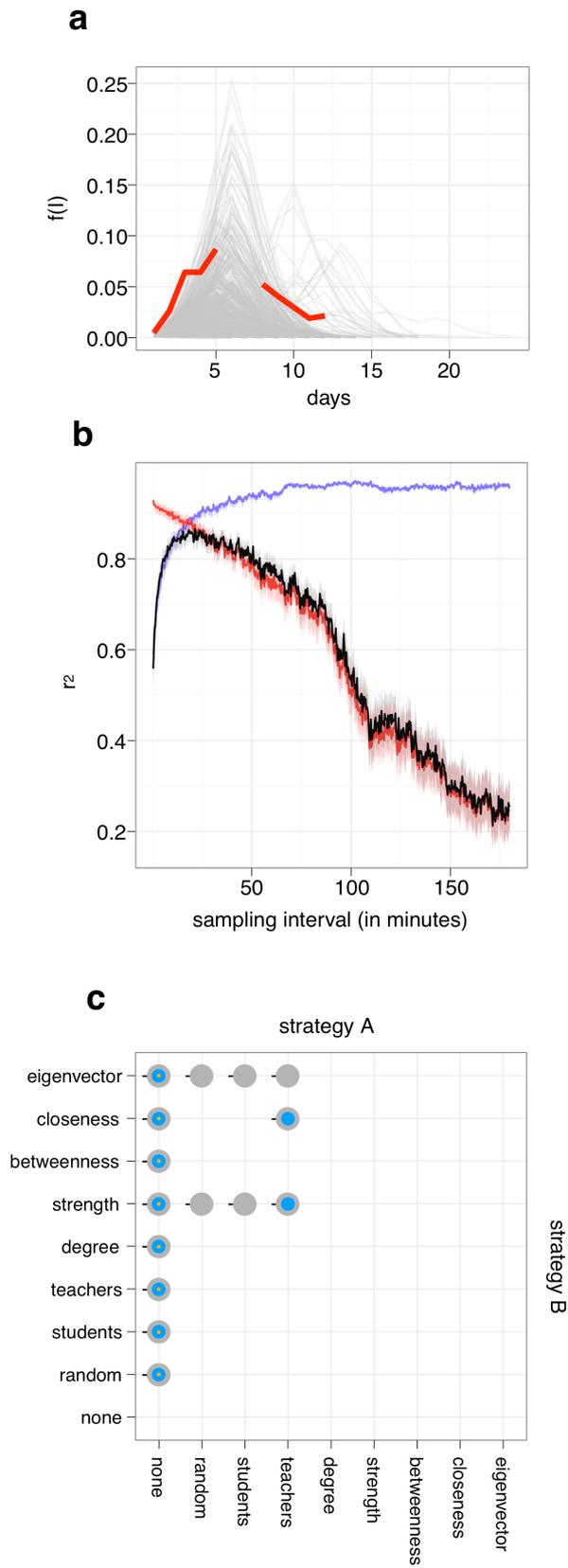

# Supplementary Material

# A High-Resolution Human Contact Network for Infectious Disease Transmission

**Marcel Salathé[1*] Maria Kazandjieva[2], Jung Woo Lee[2], Philip Levis[2], Marcus W. Feldman[1] & James H. Jones[3,4]**

1. Department of Biology, Stanford University, Stanford, CA, USA
2. Department of Computer Sciences, Stanford University, Stanford, CA, USA
2. Department of Anthropology, Stanford University, Stanford, CA, USA
3. Woods Institute for the Environment, Stanford University, Stanford, CA, USA
*E-mail: salathe@stanford.edu

**Data Collection**

Deployment Details

Motes were distributed in batches (with an average of about 11 motes) the night before the deployment and handed out to participants starting around 6am (with the vast majority of participants receiving and activating their mote upon arrival at 8am). Participants were asked to put their mote in a thin plastic pouch attached to a lanyard (provided by us) and wear the lanyard around the neck, with the mote being located in front of the chest at all times. The participants handed the motes back to us when leaving the school or at the end of the school day (the vast majority were received between 4pm and 4:30 pm). The technical details regarding code design, signal strength considerations and other issues have been described elsewhere (1), but briefly, each participant's mote was programmed to broadcast beacons at -16.9 dBm at a regular 20-second interval; the packet included the sender's local sequence number. Upon receiving a beacon, the mote checked the RSSI value (Received Signal Strength Indicator) of the packet. Note that the motes are always scanning, so no interactions of at least 20 seconds duration will be missed. If the signal strength was lower than -80 dBm, the packet was discarded (this decision was based on experimental data showing that when subjects were facing each other, packets within 3 meters had RSSI of roughly -80 or above; packets sent when one subject was facing the other person's back had a lower RSSI (1) – see also Figure S1). Otherwise, the receiver created a contact entry, consisting of the sender's ID and beacon sequence number, as well as the local mote's sequence number and the RSSI value of the packet.

TelosB motes have a 1 MB flash memory where interactions can be stored, thus eliminating the need to broadcast interactions to any other external hardware for storage. As a consequence, interactions between subjects can be captured anywhere on the campus of the school, an area of more than 45,000 square meters.

Reconstructing the full contact network required a global timestamp, relative to which all interactions between subjects occurred. Local sequence numbers in each data trace acted as relative clocks, and they could be used as offsets from one stationary mote (the "master stationary mote", located in the main cafeteria), which would be the master clock providing global time. Packets originating from this mote were transmitted at high power (-11 dBm) and were not subject to RSSI filtering at the receiver. More than 90% of mobile motes had received one or more beacons from the master stationary mote. For these mobile motes, we calculated the offset between the master and the local sequence numbers. In addition, we created a table of offsets to serve as a lookup table which included mobile motes as well as other stationary motes. To process data traces from mobile motes that did not hear directly from the master stationary mote, we used the offsets table to transitively compute a timestamp from another mote that already had its global time.

After processing the raw data, we thus obtained a list of interactions in the format

```
id1   id2  number_of_beacons  global_timestamp
```

The list contains 762,868 unique interactions between motes `id1` and `id2`, for a duration of `number_of_beacons` consecutive beacon intervals starting at time `global_timestamp`. Since beacons are broadcast every 20 seconds, the number of beacons can be used as an approximate measure of contact duration (such that duration in minutes ~ `number_of_beacons`/3).

This project was approved by the Stanford Institutional Review Board (IRB) on July 24th 2009.

**FIGURE LEGENDS**

**Figure S1**: Dependency of signal strength on distance (1, 2, 3 and 4 meters), orientation (forward or backward) and angle (0, 45, 90 and 135 degrees). The black horizontal line shows the threshold value that was chosen for the data collection. In (a), points show the average signal strength and the bars represent the standard deviation of a particular measurement. Some settings lack data because no packets were received. (A slight horizontal offset was added to the data points for visual clarity.) Panel (b) shows the spatial setting of the 4 angles and two directions, with reference to the main mote.

**FIGURES**

**Figure S1**

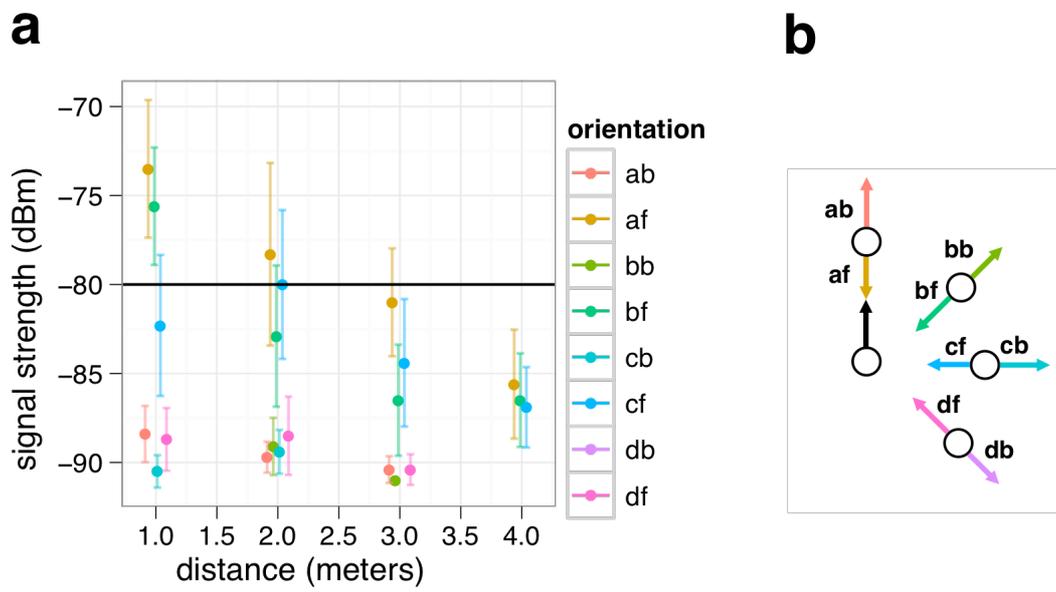